\documentclass[a4paper]{jpconf}
\usepackage{graphicx}
\usepackage{subfigure}
\input{aliases}
\usepackage{hyperref}    
\usepackage[all]{hypcap} 

\begin{document}
\title{Hadron Spectroscopy, exotics and $B_c^+$ physics at LHCb}

\author{Biplab Dey on behalf of the LHCb Collaboration}

\address{Sezione INFN di Milano, Milano, Italy}

\ead{biplab.dey@cern.ch}

\begin{abstract}
The LHCb experiment is designed to study properties and decays of heavy flavored hadrons produced from $pp$ collisions at the LHC. During Run~1, it has recorded the world's largest data sample of beauty and charm hadrons, enabling precision spectroscopy studies of such particles. Several important results obtained by LHCb, such as the discovery of the first pentaquark states and the first unambiguous determination of the $Z_c(4430)^-$ as an exotic state, have dramatically increased the interest on spectroscopy of heavy hadrons. An overview of the latest LHCb results on the subject, including the discovery of four strange exotic states decaying as $X \to \jpsi \phi$, is presented. LHCb has also made significant contributions to the field of $B_c^+$ physics, the lowest bound state of the heavy flavor $\bbar$ and $c$ quarks. A synopsis of the the latest results is given.
\end{abstract}

\section{Introduction}
\label{sec:intro}

The quark model enunciated by Gell-Mann \etal~\cite{GellMann:1964nj} predicts, besides conventional $q \bar{q}$ mesons and $qqq$ baryons, any other SU(3) color-neutral combination of quarks and gluons such as $gg$ glueballs, $q\bar{q}g$ hybrids, $q \bar{q}q \bar{q}$ tetraquarks, $qqqqq$ pentaquarks, and so on. Experimentally, such ``exotic'' states have been hard to find, especially in the light quark sector. The situation changed in 2003, after the discovery of the narrow exotic tetraquark state, $X(3872)$, by the Belle collaboration~\cite{Choi:2003ue}. Since then, several other states with exotic configurations have been seen, interestingly, all involving heavy quark systems. With the first unambiguous spin-parity assignments of the $X(3872)$~\cite{Aaij:2013zoa} and $Z_c(4430)^-$~\cite{Aaij:2014jqa} states, the discovery of two pentaquark states~\cite{Aaij:2015tga}, among other results, the LHCb experiment has been contributing heavily towards our understanding of these particles. In this talk we touch upon some recent results from LHCb in exotic hadron spectroscopy, as well as in the rapidly evolving field of the doubly heavy-flavored $B_c^+$ meson.

\section{$\jpsi\phi$ exotic states in $\Bp \to \jpsi \phi \Kp$}
\label{sec:jpsiphik}

The $X(4140)$ state, first claimed by the CDF collaboration in 2008~\cite{Aaltonen:2009tz} as a narrow, near-threshold peak in the $\jpsi\phi$ invariant mass in the decay $\Bp \to \jpsi \phi \Kp$, has generated both considerable interest as well as confusion. Results from several experiments~\cite{Aaltonen:2009tz,Shen:2009vs,Aaij:2012pz,Chatrchyan:2013dma,Abazov:2013xda,Lees:2014lra} have seen disagreements on the properties of the $X(4140)$ and higher lying states. Theoretical interpretations range from tetraquark, molecular or hybrid states, or a re-scattering phenomena, including the so-called ``cusp effect''~\cite{Swanson:2015bsa} at the $D^\pm_s D^{\ast \mp}_s$ production threshold. An earlier LHCb analysis based on a data sample corresponding to $0.37~\invfb$ of integrated luminosity at $\sqrts=7$~TeV found no evidence of the $X(4140)$ state. With a larger data sample corresponding to the entire Run~I dataset, the latest LHCb results~\cite{Aaij:2016nsc,Aaij:2016iza} incorporate an amplitude analysis of the complete decay chain $\Bp \to \jpsi(\to \mup \mun) \phi(\to \Kp \Km) \Kp$. The estimated signal yield is $N_{\rm sig} =4289\pm 151$, and the estimated background fraction in the signal region is $(23\pm6)\%$. This constitute the largest and cleanest world dataset for this decay mode.

For the amplitude analysis, there are three relevant production amplitudes that can contribute: $\Bp \to \jpsi K^{\ast +} (\to \phi K^+)$, $\Bp \to X (\to \jpsi \phi) \Kp$, and $\Bp \to Z^+_c (\to \jpsi K^+)\phi$.
Each of the three decay amplitudes are constructed in its corresponding helicity basis. The spin-projections of all intermediate states are summed over coherently, while the spin-projections of the final state muons are summed incoherently. The lineshapes of resonant structures are taken as relativistic Breit-Wigners with mass-dependent widths, folded with the corresponding angular-momentum barrier factor. For non-resonant components, the lineshape is taken just as the angular-momentum barrier factor. The transition matrix element depends on six kinematic variables describing the full decay chain, and the angular analysis fits to the joint six-dimensional differential rate.

\begin{figure}[h]
\centering
\subfigure[]{
\centering
\includegraphics[width=0.48\textwidth]{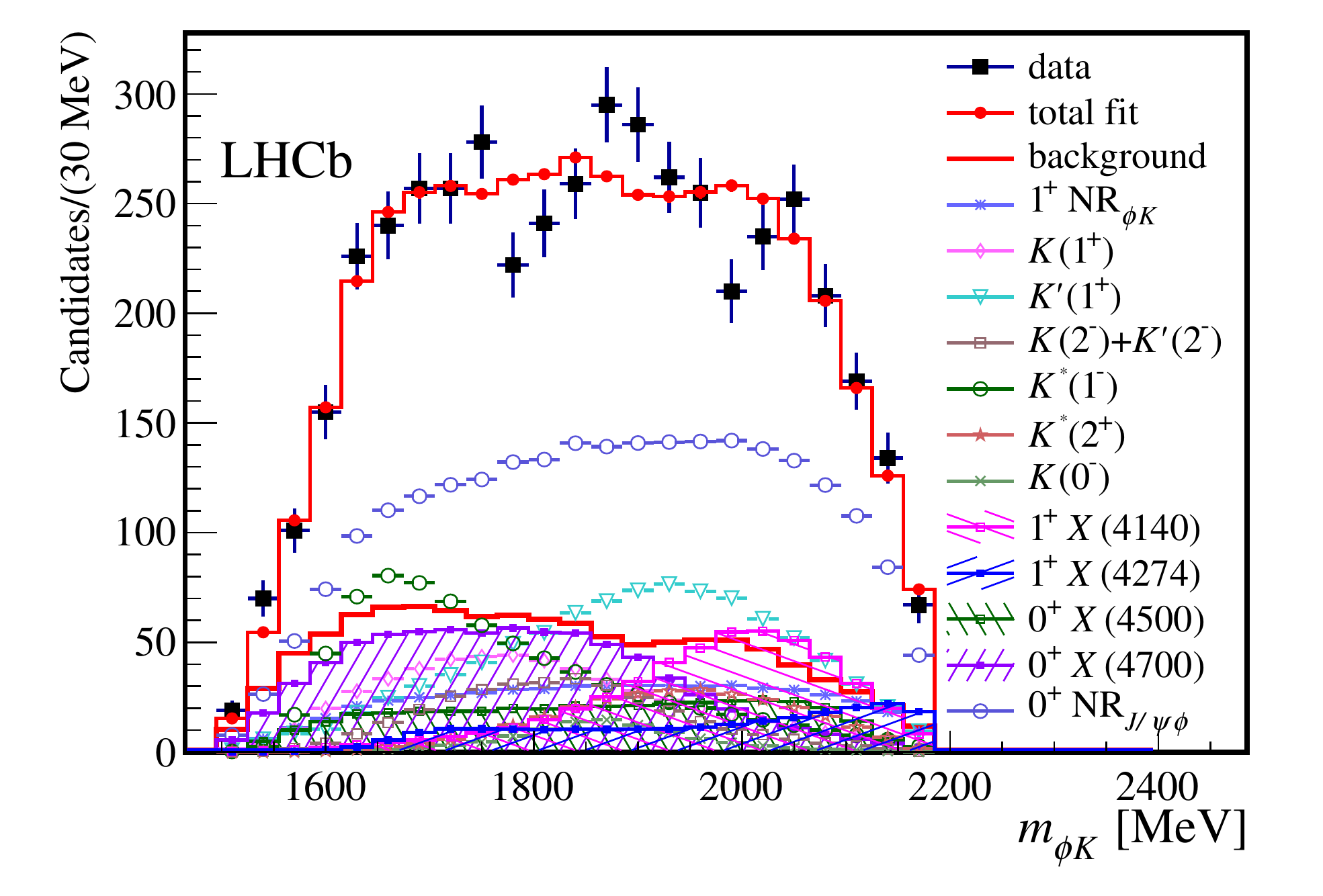}
}
\subfigure[]{
\centering
\includegraphics[width=0.48\textwidth]{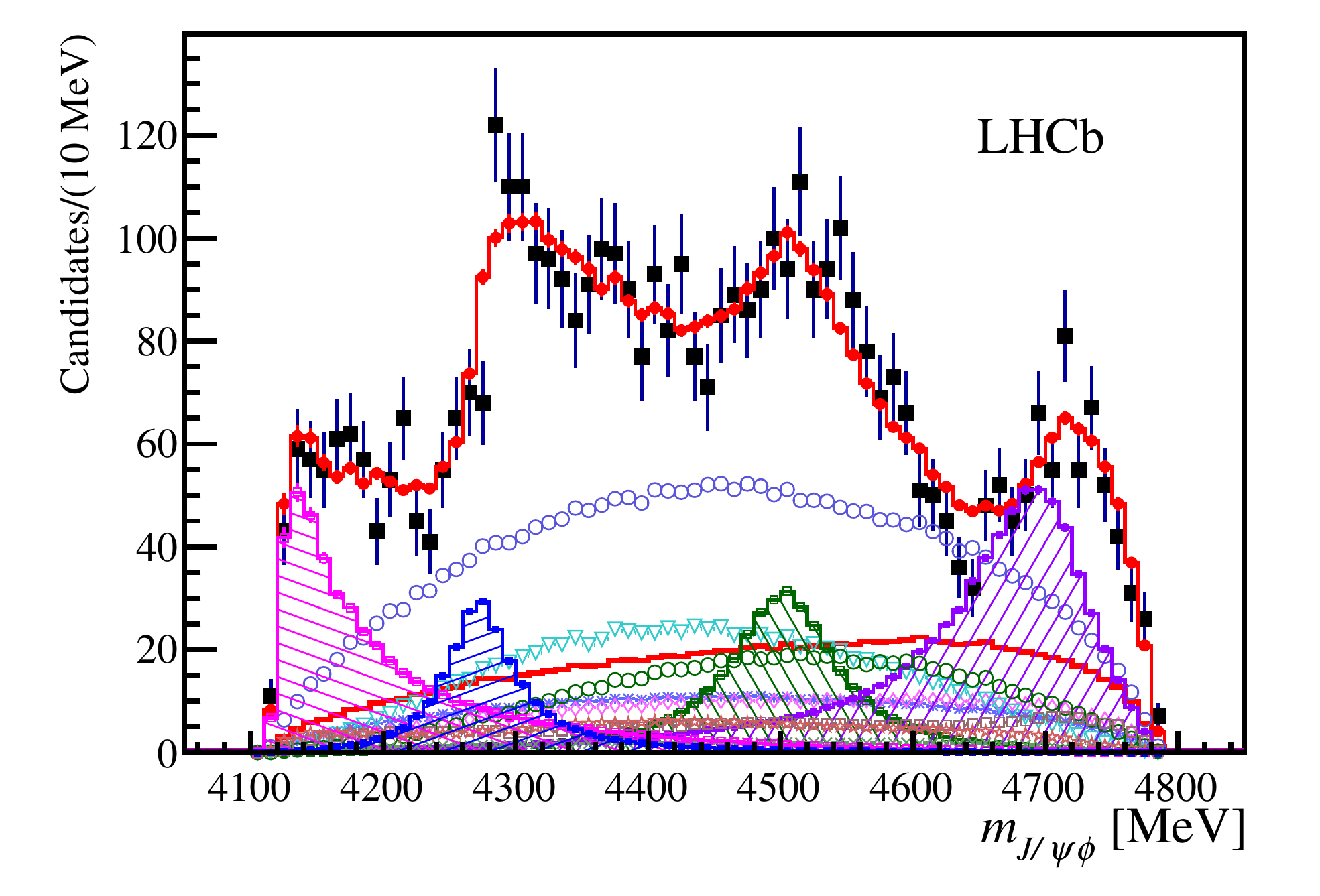}
}
\caption{Results of the amplitude fit projected on to variables (a) $m(\phi K)$ and (b) $m(\jpsi \phi)$.} 
\label{fig:jpsiphik_fit_results}
\end{figure}

Figure~\ref{fig:jpsiphik_fit_results} shows the results of the default amplitude fit, projected onto the ``$K^\ast$'' and ``$X$'' invariant mass variables (the exotic $Z_c$ contributions are found to not dominate). Although the overall $m(\phi K)$ distribution in Fig.~\ref{fig:jpsiphik_fit_results}a appears flat, a rich spectrum of $K^{\ast +}\to \phi \Kp$ resonances are found in the angular distributions. 
The near-threshold $\jpsi \phi$ structure in Fig.~\ref{fig:jpsiphik_fit_results}b is found to be consistent with previous world data on $X(4140)$ in the mass, while the width is found to be larger. A nearby second state consistent with the $X(4272)$ found by the CDF collaboration~\cite{Aaltonen:2011at} is seen, and both states prefer spin-parity assignments $J^P=1^+$. The higher $m(\jpsi \phi)$ region is also found to require two $J^P = 0^+$ resonances, $X(4500)$ and $X(4700)$. The significances of all four exotic states are found to exceed $5~\sigma$.


\section{Reconfirmation of the pentaquarks}

\subsection{Pentaquarks in $\Lb \to \jpsi p \pim$}
\label{sec:jpsippi}

\begin{figure}[h]
\centering
\subfigure[]{
\centering
\includegraphics[width=1.95in]{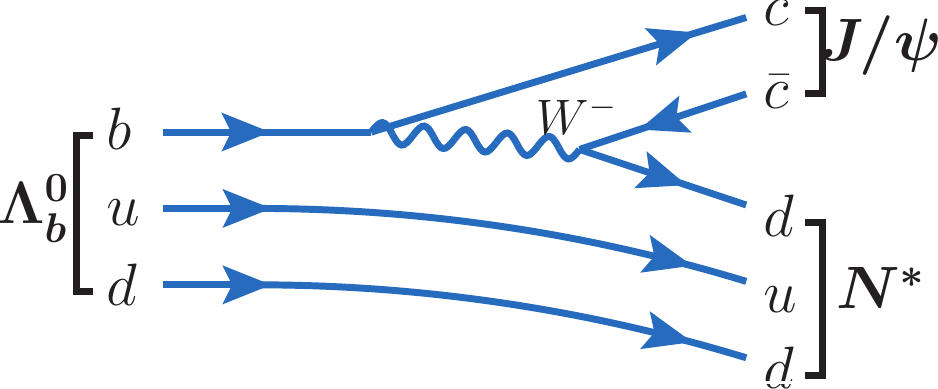}
}
\subfigure[]{
\centering
\includegraphics[width=1.95in]{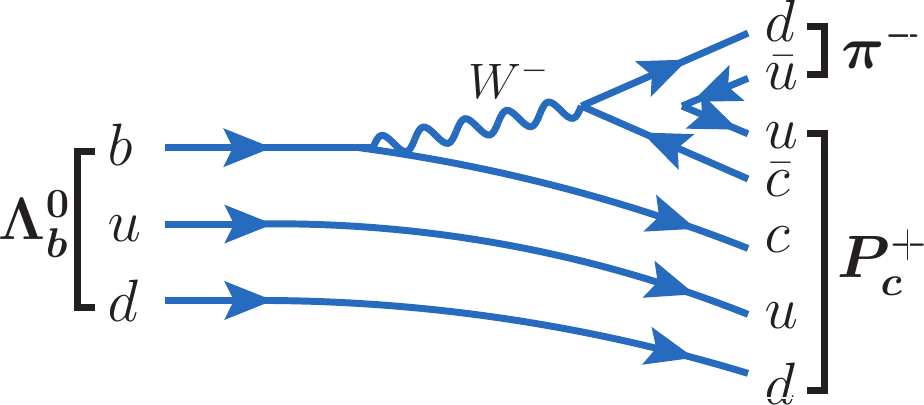}
}
\subfigure[]{
\centering
\includegraphics[width=1.95in]{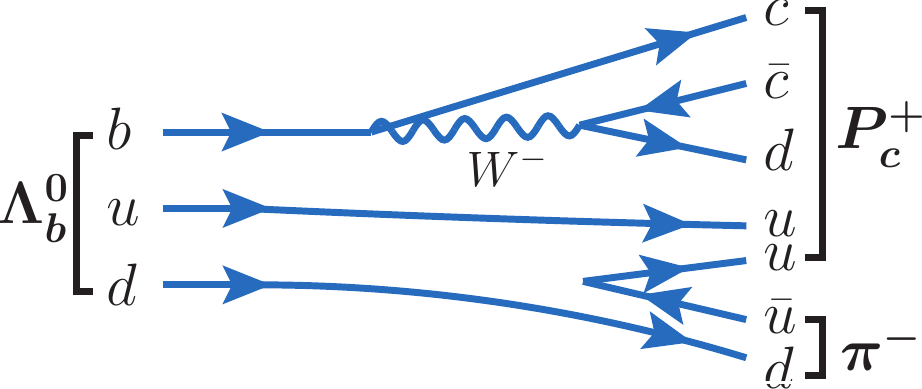}
}
\caption{Feynman diagrams for the decay $\Lb \to \jpsi p \pim$: (a) $N^\ast$ resonances (b) ``external'' $W^{\ast -}$ exchange, and (c) ``internal'' $W^{\ast -}$ exchange.}
\label{fig:jpsippi_feyn}
\end{figure}

After the 2015 discovery of two pentaquark states $P_c(4380)^+$ and $P_c(4450)^+$ in the mode $\Lb \to P_c^+(\to \jpsi p)K^-$ by the LHCb collaboration~\cite{Aaij:2015tga}, one question that naturally arises is if these states occur in other decay modes as well. Analysing the same Run~I dataset, but now in the Cabibbo suppressed mode, $\Lb \to \jpsi p \pim$, the data is found to be consistent with the two pentaquarks coupling via the $\Lb \to P_c^+ \pim$ mode~\cite{Aaij:2016ymb}. Figures~\ref{fig:jpsippi_feyn}a and~\ref{fig:jpsippi_feyn}b show the Feynman diagrams in $\Lb\to \jpsi p \pim$, analogous to those in the $\Lb\to \jpsi pK^-$~\cite{Aaij:2015tga}, but with the $s$-quark now replaced by a $d$-quark. In addition, for the $p\pim$ mode, the third diagram in Fig.~\ref{fig:jpsippi_feyn}c contributes as well, which can potentially result in the ratio of the branching fractions $R_{\pi/K} = \mathcal{B}(\Lb\to P^+_c\pi^-)/\mathcal{B}(\Lb\to P^+_c K^-)$ to deviate from that expected from Cabibbo suppression~\cite{Cheng:2015cca}. 

\begin{figure}[h]
\centering
\subfigure[]{
\centering
\includegraphics[width=2.6in]{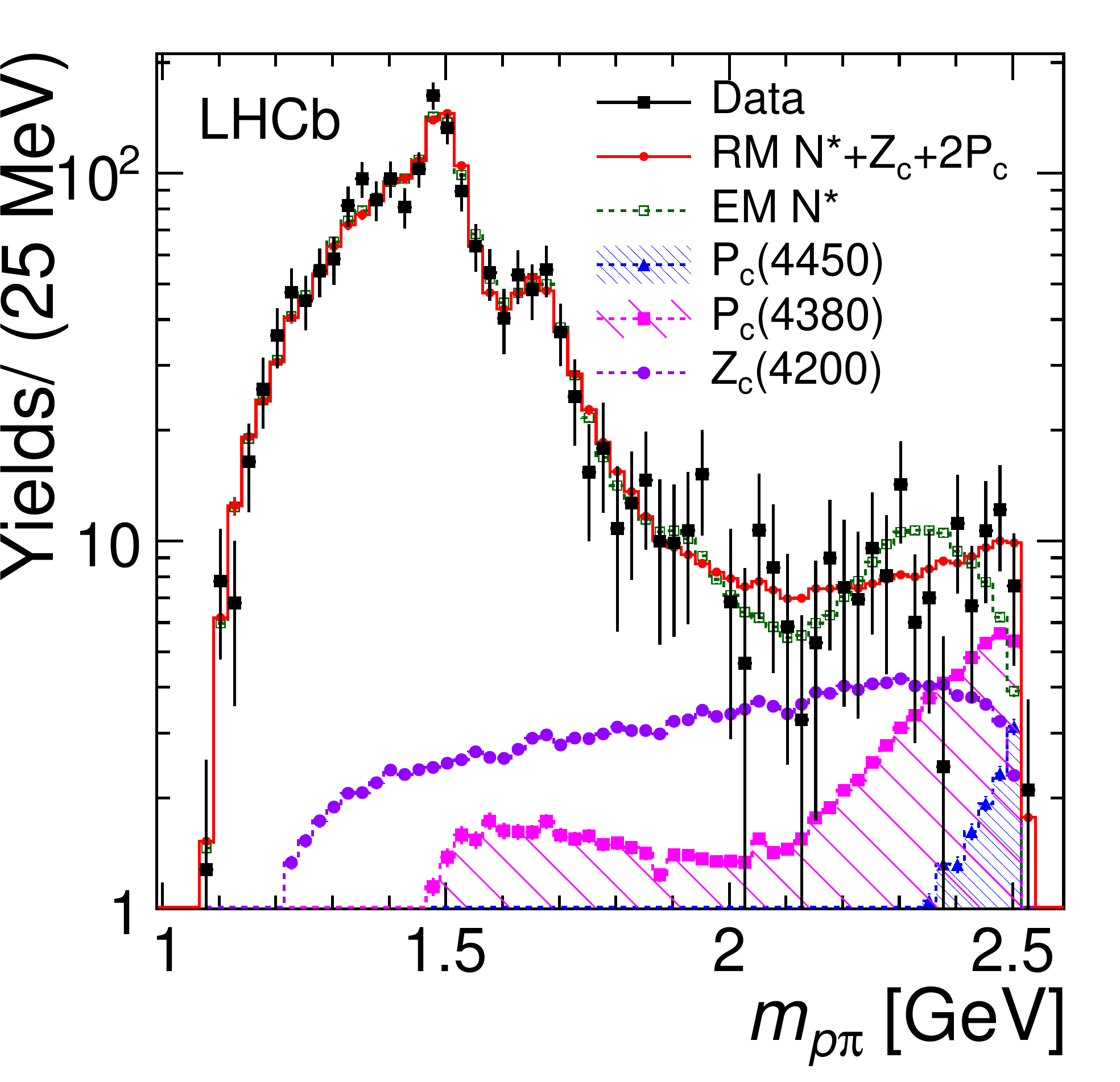}
}
\subfigure[]{
\centering
\includegraphics[width=2.6in]{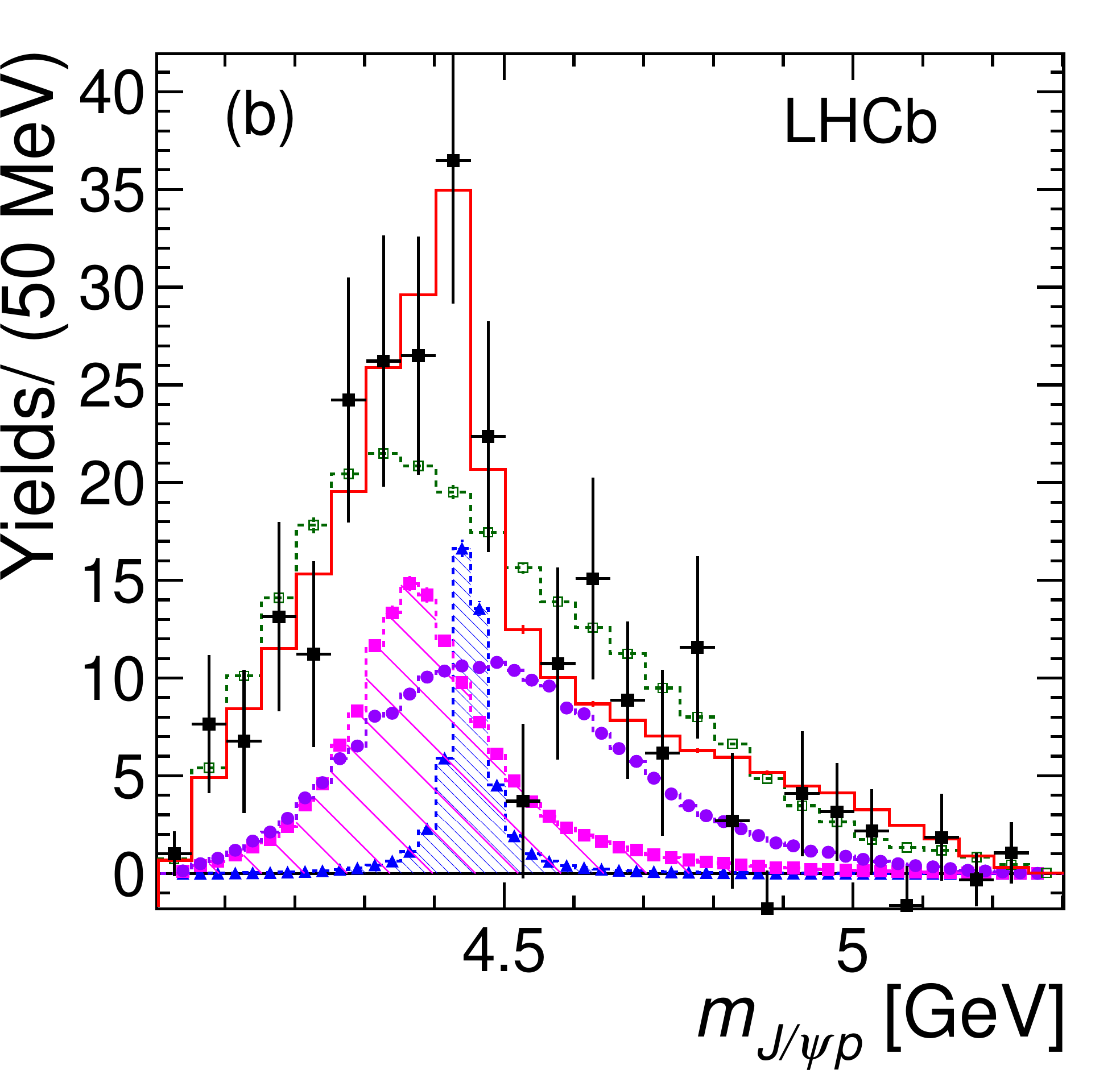}
}
\caption{Projections of the background-subtracted data and the amplitude fit results in $\Lb \to \jpsi p \pim$ projected on to: (a) $m(p\pim)$, and (b) $m(\jpsi p)$ with $m(p\pim)> 1.8$~GeV.}
\label{fig:jpsippi_ampfits}
\end{figure}

Since $p \pim$ is a Cabibbo-suppressed mode, compared to $p \Km$, the statistics is around 15 times lesser ($N_{\rm sig} = 1885\pm 50$), with thrice the percentage of background. The dominant contributions to the decay amplitude come from $\Lb \to \jpsi N^\ast(\to p \pim)$. $N^\ast$ resonances are modeled as relativistic Breit-Wigners, except the $N(1535)$, which is modeled by a Flatte lineshape, since, besides coupling to $p\pim$, it couples to the near-threshold $n\eta$ mode as well. The last component contributing is $\Lb \to Z_c^-(\to \jpsi \pim) p$, where $Z_c^-$ is an exotic contribution, such as the $Z_c(4200)^-$ tetraquark candidate claimed by the Belle collaboration~\cite{Chilikin:2014bkk}. Since the Cabibbo suppressed dataset is limited in statistics, the properties of the the $P_c$ and $Z_c$ exotic candidates are kept as in previous world data and a full amplitude fit is performed, following the formalism in the original pentaquark analysis~\cite{Aaij:2015tga}. Two different models of the $N^\ast$ contributions are included: ``reduced'' (RM), and ``extended'' (EM), the latter including a wider set of resonances and allowed couplings. Figure~\ref{fig:jpsippi_ampfits} shows the background-subtracted data and fits results projected on to the variables $m(p\pim)$ and $m(\jpsi p)$. The pentaquark contributions are more prominent at higher $m(p\pim)$ masses, as visible in Fig.~\ref{fig:jpsippi_ampfits}b.

The significance for exotic contributions (the $P_c$'s and the $Z_c(4200)^-$) is found to be $3.1~\sigma$ and $R_{\pi/K}= 0.05$, and 0.033, for $P_c(4380)^+$ and $P_c(4450)^+$, respectively, consistent with expectations from Cabibbo suppression~\cite{Cheng:2015cca}, pointing to a negligible contribution from the diagram in Fig.~\ref{fig:jpsippi_feyn}c.

\subsection{Model-independent confirmation of exotic activity in $\Lb \to \jpsi p \Km$}
\label{sec:pc}

\begin{figure}[h]
\centering
\subfigure[]{
\centering
\includegraphics[width=0.38\textwidth]{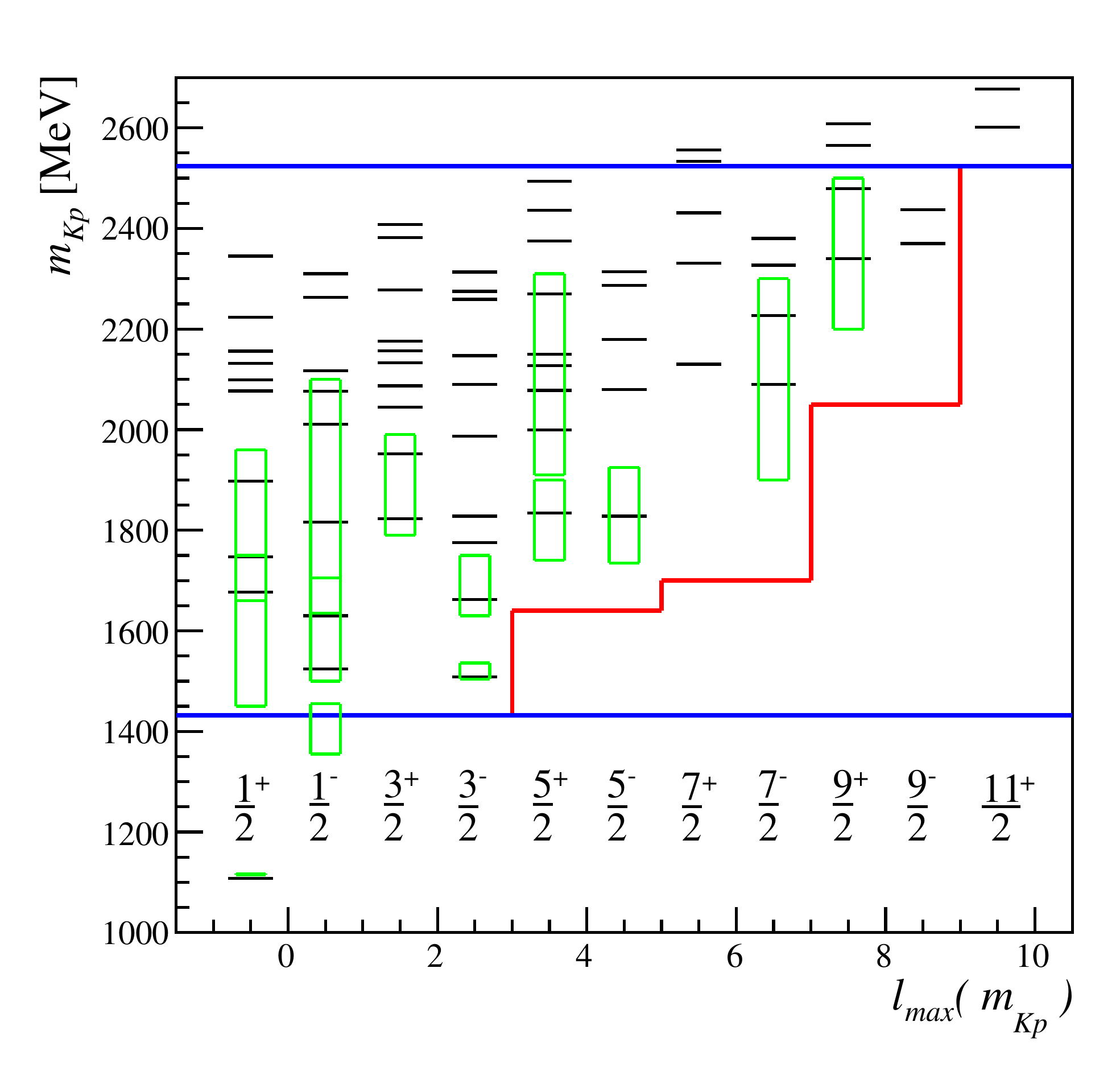}
}
\subfigure[]{
\centering
\includegraphics[width=0.51\textwidth]{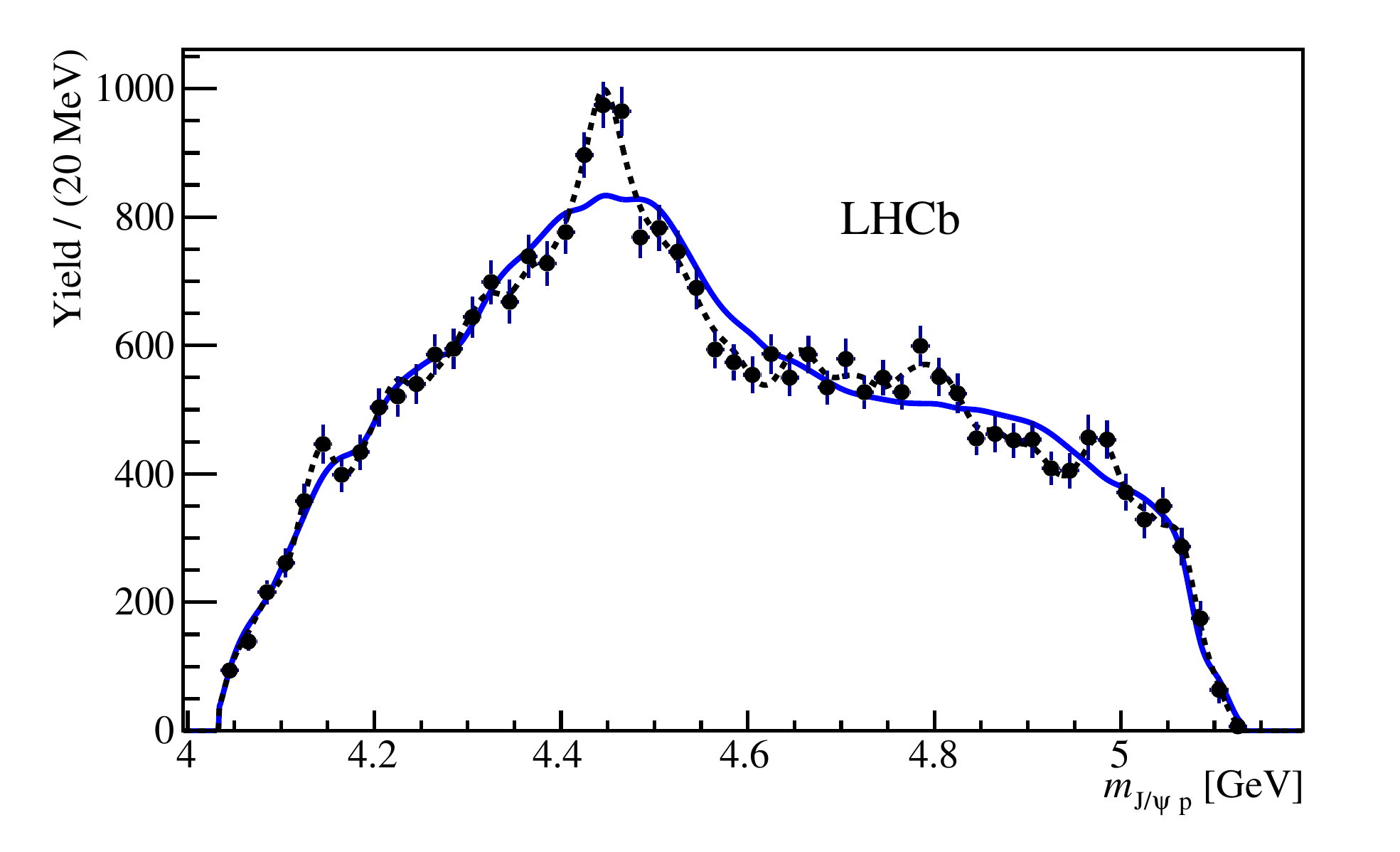}
}
\caption{Model-independent exotic evidence in $\Lb \to \jpsi p \Km$: (a) at a given $m(p\Km)$, the range of expected spin-$J$ physical $\Lambda^\ast_J$ resonances, where $l_{\rm max}=2J_{\rm max}+1$ and the green bands are the experimentally observed states; (b) comparison between the background subtracted and efficiency corrected data (black markers), predictions from a non-exotic ``$\Lambda^\ast_J$-only'' model (blue curve) and exotic ``$l_{\rm max}=31$'' predictions (broken lines).}
\label{fig:pc}
\end{figure}

One of the important sources of systematic uncertainties in the exotic states searches is the often poorly known spectrum of conventional hadronic resonances. For pentaquarks in $\Lb \to \jpsi p \Km$~\cite{Aaij:2015tga}, these comprise the exited $\Lambda^\ast_J\to p\Km$ states of spin-$J$. To circumvent this problem, the LHCb adopted a model-independent method~\cite{Aaij:2016phn} first proposed by the BaBar collaboration in the context of $Z_c(4430)^-$ searches~\cite{Aubert:2008aa}. The underlying principle is that if the highest spin of the the contributing $\Lambda^\ast_J$ states is $J_{\rm max}$, then the maximal power of $x=\cos \theta_{pK}$ in the differential decay rate is given by the Legendre polynomial, $P_{l_{\rm max}}(x)$, of order $l_{\rm max}=2J_{\rm max}+1$. Here $\theta_{pK}$ is the helicity angle of the daughter kaon in the mother $\Lambda^\ast_J$ rest-frame. Typically, physical $\Lambda^\ast_J$ resonances occurring at a given $m(pK)$ mass will have a limited ranges of $l_{\rm max}$ as depicted in Fig.~\ref{fig:pc}a: resonances of higher spins occur at higher masses, as demarcated by the red and blue lines.

On the other hand, possible exotic states in the $\jpsi p$ or $\jpsi \Km$ systems ``reflect'' on the entire spectrum of spin-$J$ $p\Km$ states in the 3-body Dalitz plane $\Lb \to \jpsi p \Km$. At a given $m(p\Km)$ invariant mass, the presence of such unphysically large spin-$J$ components, corresponding to large $l_{\rm max}$, beyond the limits in Fig.~\ref{fig:pc}a, signals the present of exotic activity. Further, since the Legendre polynomials constitute a complete basis of orthonormal functions, the $l_{\rm max}$ component in the $\cos \theta_{pK}$ distribution can be extracted via a counting experiment, weighting each event by the corresponding $P_{l_{\rm max}}(\cos \theta_{pK})$ function. The analysis uses the same Run~I dataset and selection as in the earlier LHCb pentaquark paper.~\cite{Aaij:2015tga}. Figure~\ref{fig:pc}b shows the efficiency-corrected and background-subtracted distribution in $m(\jpsi p)$, compared with predictions from an ``exotic'' model with large $l_{\rm max}=31$ (shown by the broken lines) that can accommodate reflections from exotics, compared to a non-exotic ``$\Lambda^\ast_J$-only'' model (shown by the blue curve), with $l_{\rm max}$ as a function of $m(p\Km)$ given by Fig.~\ref{fig:pc}a. The ``$\Lambda^\ast_J$-only'' model clearly does not describe the data, which is a strong pointer towards the presence of exotic activity. Based on a likelihood ratio study using pseudoexperiments, between the ``$\Lambda^\ast_J$-only'' and ``exotic'' models, LHCb was also able to place a numeric estimate of the exotic significance at above $9~\sigma$.

We underscore the point that the efficacy of this method is its independence from requiring any detailed knowledge of the complicated $\Lambda^\ast_J$ spectrum.

\section{Non-confirmation of the $X(5568)$ tetraquark}
\label{sec:x5568}

The D0 collaboration has recently claimed a 5.1~$\sigma$ evidence for a novel ``4-flavored'' ($\bar{b}su\bar{d}$) exotic tetraquark state $X(5568)^\pm\to \Bs\pipm$, using $p\bar{p}$ collisions at $\sqrt{s}=1.97$~TeV. Further, the relative production rate between the $X(5568)$ and $\Bs$, multiplied by the $X\to \Bs\pipm$ branching fraction has been claimed to be $\rho_X^{\rm D0}\sim 8.6\%$.

\begin{figure}[h]
\centering
\subfigure[]{
\centering
\includegraphics[width=0.48\textwidth]{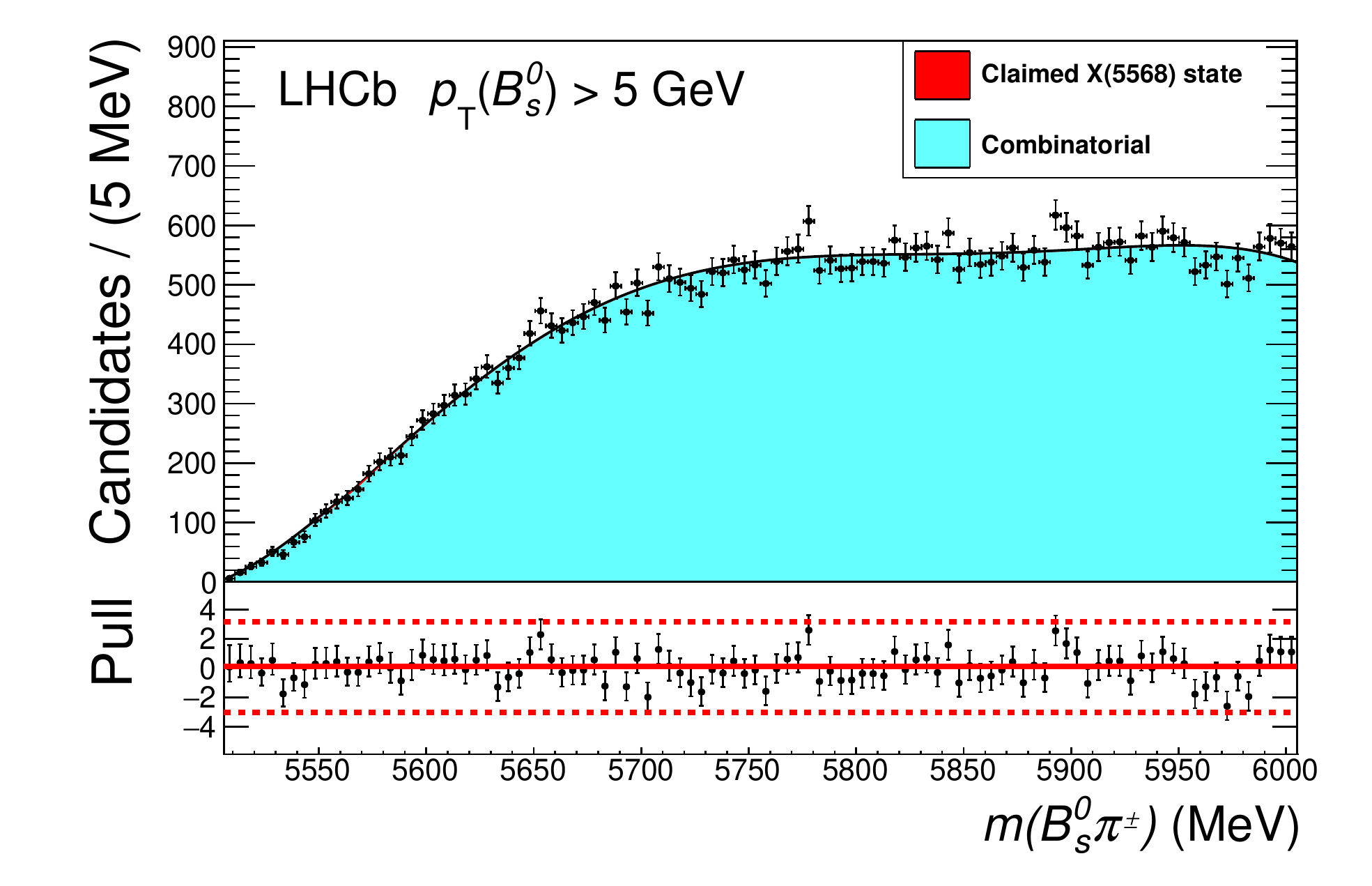}
}
\subfigure[]{
\centering
\includegraphics[width=0.48\textwidth]{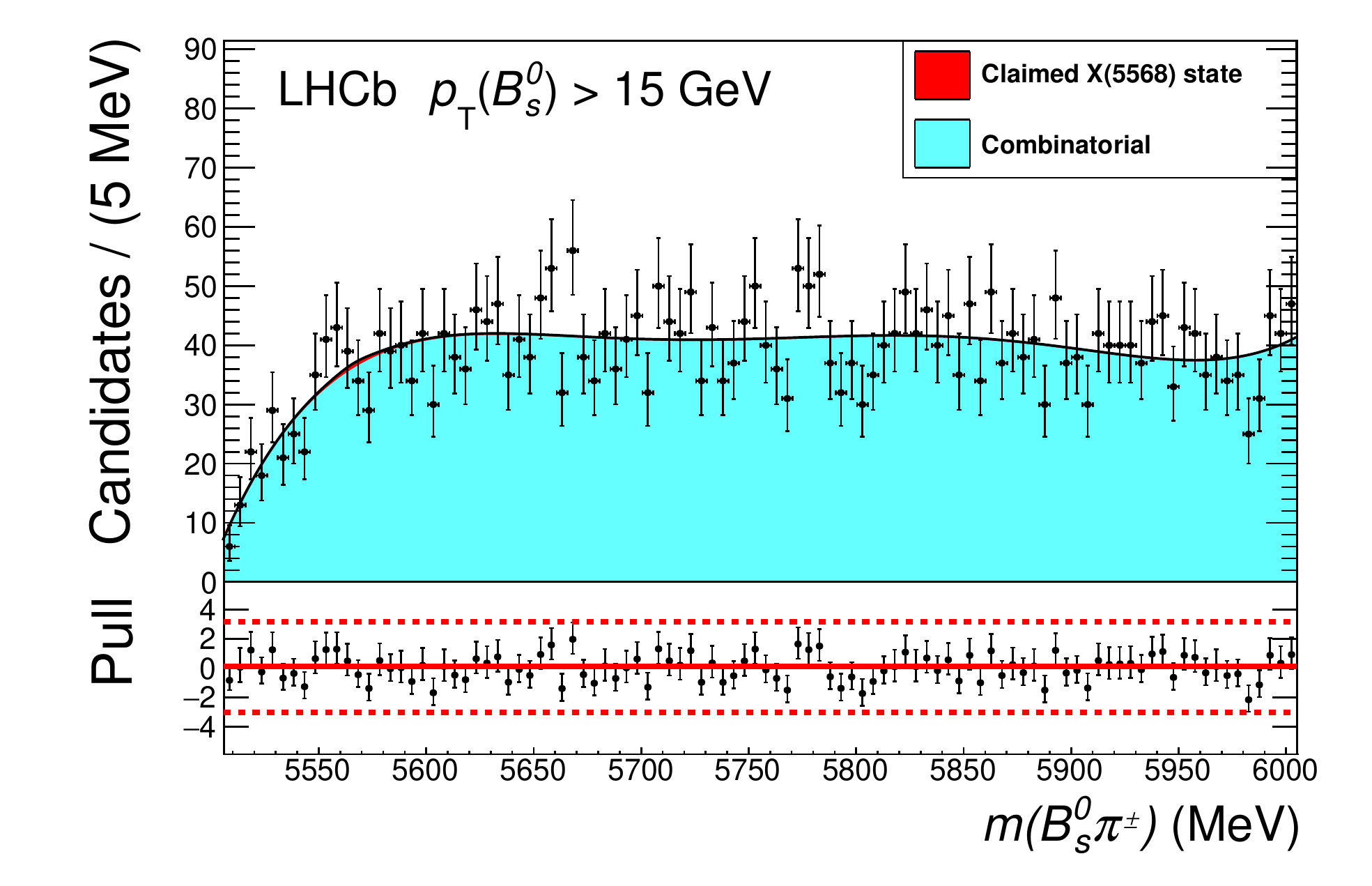}
}
\caption{Fits to the LHCb $\Bs\pipm$ spectrum~\cite{Aaij:2016iev} showing no significant excess at the $X(5568)$ mass for $\pt(\Bs)$ greater than (a) 5~GeV and (b) 15~GeV.}
\label{fig:x5568_lhcb}
\end{figure}

LHCb has searched for the $X(5568)$ decaying to $\Bs\pipm$ with the full Run~I dataset~\cite{Aaij:2016iev}, employing the modes $\Bs\to D_s^- \pip$ and $\Bs \to \jpsi \phi$, with $D^-_s\to \Kp\Km\pim$, $\jpsi \to \mup \mun$ and $\phi \to \Kp \Km$. The overall signal yield is around 110,000 $B_s$ decays, roughly twenty times larger than in D0, and with a higher purity. The selection requirements for the $\Bs$ and the companion $\pipm$ follow those in previous well-understood LHCb analyses. To facilitate a clean extraction of the $\Bs$ candidates, its transverse momentum, $\pt$, is required to be greater than 5, 10, or 15~GeV. No significant $X(5568)$ is seen for either of the three $\pt(\Bs)$ choices, as shown in Fig.~\ref{fig:x5568_lhcb} and upper limits of the order $\rho_X^{\rm LHCb}\sim 2\%$ at the $95\%$ confidence level, are placed.

We also note that preliminary results from the CMS collaboration~\cite{CMS:2016fvl} also indicate toward non-observation of any resonant-like structure around the purported $X(5568)$ mass. Results from the ATLAS and CDF collaborations are still awaited.

\section{$B_c^+$ physics at LHCb}

The $B_c^+$ ($\bar{b}c$) meson is unique in being the only established hadronic state composed of two different heavy flavor quarks. Unlike the $b\bar{b}$ or the $c\bar{c}$ onia states, due to its flavor quantum numbers $B=-C=\pm1$, the $B_c^+$ can not decay strongly, but only weakly, either via $\bar{b}\to \bar{c} W^+$, $c\to s W^+$, or the so-called weak-annihilation (WA) process $\bar{b}c\to W^+$. The latter can potentially receive contributions New Physics, such as charged Higgs exchange at the tree-level. Further, compared to $B^+$ decays, the WA process in $B_c^+$ decays is Cabibbo favored by the factor $|\vcb/\vub|^2\sim 100$. LHCb has performed several measurements involving the $B_c^+$, including its decay modes with charmonium final states $\jpsi 3\pi$, $\jpsi\Kp$, $\psi(2S)\pip$, $\jpsi D_s^{(\ast)+}$, $\jpsi\Kp\Km\pip$, $\jpsi3\pip2\pim$ that derive from the the Cabibbo suppressed $\bbar \to \bar{c}$ transition, as well $\B^+_c\to B_s^0\pip$, involving the Cabibbo favored $c \to s$ transition. The most precise measurements of its mass~\cite{Aaij:2014asa} and lifetime~\cite{Aaij:2014gka} come from LHCb as well and are consistent with expectation from Lattice QCD and heavy quark effective theory.

More recently, there has been emphases on WA studies via charmless $B_c^+ \to h^+ h^- h'$ decays such as $B_c^+\to \{KKK, \pi\pi\pi,KK\pi,p \bar{p}K,p \bar{p}\pi\}$~\cite{Aaij:2016xxs,Aaij:2016xas}. Theory predictions for the WA processes exist only for the quasi two-body modes, with branching fractions ranging in $10^{-6}$ to $10^{-8}$~\cite{DescotesGenon:2009ja}. Since the $B_c^+$ has a mass of around 6274~MeV, decays to three light hadrons result in large available phase-space and the formation of intermediate $D^0$, $B^0_{(s)}$ and $\psi^{(')}$ resonances also contribute. Using the full Run~I dataset, LHCb has searched for the rare decay $B_c^+\to p \bar{p}\pip$~\cite{Aaij:2016xxs} and also measured $R_p\equiv \frac{f_c}{f_u}\times \mathcal{B}(B_c^+ \to p\bar{p}\pip)$, where $f_c(f_u)$ is the fragmentation fraction of a $b$ quark into $B_c^+$ ($B^+$). In the $\jpsi$ veto region, $m(p \bar{p})<2.85$~GeV, no signal was detected and an upper limit at the $95\%$ confidence level is set as $R_p < 3.6\times 10^{-8}$. The latest LHCb results~\cite{Aaij:2016xas} on the $B_c^+\to \Kp \Km \pip$ mode has seen some hints of WA, but more data is needed.

\section{Conclusions}

During the Run~I phase of LHC running, the LHCb experiment has made several important strides in exotic hadron spectroscopy searches and $B_c^+$ physics. The original pentaquark discovery has been reconfirmed by two further measurements and evidence for four possible new strange exotics have been seen in the $\jpsi \phi$ spectrum, while the $X(5558)$ tetraquark state claimed by the DO collaboration stands unconfirmed at the LHC, as yet. With a significant production and acceptance rate of $B_c^+$ mesons inside LHCb detector, several precision or first measurements of its mass, lifetime and decay modes have been made. The interesting problem of whether the weak annihilation process $\bbar c\to W^+$ in $B_c^+$ 3-body charmless decays, can be enhanced from New Physics contributions has also being probed. Studies are also ongoing on the spectroscopy of $B_c^{+\ast}$ states. With a five fold increase in statistics expected the ongoing Run~II data-taking period ends in 2018, further detailed studies of these questions are expected.

\section*{References}
\bibliography{biblio}

\providecommand{\newblock}{}
\begin{thebibliography}{10}
\expandafter\ifx\csname url\endcsname\relax
  \def\url#1{{\tt #1}}\fi
\expandafter\ifx\csname urlprefix\endcsname\relax\def\urlprefix{URL }\fi
\providecommand{\eprint}[2][]{\url{#2}}

\bibitem{GellMann:1964nj}
Gell-Mann M 1964 {\em Phys. Lett.\/} {\bf 8} 214--215

\bibitem{Choi:2003ue}
Choi S~K {\em et~al.\/} (Belle) 2003 {\em Phys. Rev. Lett.\/} {\bf 91} 262001
  (\textit{Preprint} \eprint{hep-ex/0309032})

\bibitem{Aaij:2013zoa}
Aaij R {\em et~al.\/} (LHCb) 2013 {\em Phys. Rev. Lett.\/} {\bf 110} 222001
  (\textit{Preprint} \eprint{1302.6269})

\bibitem{Aaij:2014jqa}
Aaij R {\em et~al.\/} (LHCb) 2014 {\em Phys. Rev. Lett.\/} {\bf 112} 222002
  (\textit{Preprint} \eprint{1404.1903})

\bibitem{Aaij:2015tga}
Aaij R {\em et~al.\/} (LHCb) 2015 {\em Phys. Rev. Lett.\/} {\bf 115} 072001
  (\textit{Preprint} \eprint{1507.03414})

\bibitem{Aaltonen:2009tz}
Aaltonen T {\em et~al.\/} (CDF) 2009 {\em Phys. Rev. Lett.\/} {\bf 102} 242002
  (\textit{Preprint} \eprint{0903.2229})

\bibitem{Shen:2009vs}
Shen C~P {\em et~al.\/} (Belle) 2010 {\em Phys. Rev. Lett.\/} {\bf 104} 112004
  (\textit{Preprint} \eprint{0912.2383})

\bibitem{Aaij:2012pz}
Aaij R {\em et~al.\/} (LHCb) 2012 {\em Phys. Rev.\/} {\bf D85} 091103
  (\textit{Preprint} \eprint{1202.5087})

\bibitem{Chatrchyan:2013dma}
Chatrchyan S {\em et~al.\/} (CMS) 2014 {\em Phys. Lett.\/} {\bf B734} 261--281
  (\textit{Preprint} \eprint{1309.6920})

\bibitem{Abazov:2013xda}
Abazov V~M {\em et~al.\/} (D0) 2014 {\em Phys. Rev.\/} {\bf D89} 012004
  (\textit{Preprint} \eprint{1309.6580})

\bibitem{Lees:2014lra}
Lees J~P {\em et~al.\/} (BaBar) 2015 {\em Phys. Rev.\/} {\bf D91} 012003
  (\textit{Preprint} \eprint{1407.7244})

\bibitem{Swanson:2015bsa}
Swanson E~S 2016 {\em Int. J. Mod. Phys.\/} {\bf E25} 1642010
  (\textit{Preprint} \eprint{1504.07952})

\bibitem{Aaij:2016nsc}
Aaij R {\em et~al.\/} (LHCb) 2016  (\textit{Preprint} \eprint{1606.07898})

\bibitem{Aaij:2016iza}
Aaij R {\em et~al.\/} (LHCb) 2016  (\textit{Preprint} \eprint{1606.07895})

\bibitem{Aaltonen:2011at}
Aaltonen T {\em et~al.\/} (CDF) 2011  (\textit{Preprint} \eprint{1101.6058})

\bibitem{Aaij:2016ymb}
Aaij R {\em et~al.\/} (LHCb) 2016 {\em Phys. Rev. Lett.\/} {\bf 117} 082003
  (\textit{Preprint} \eprint{1606.06999})

\bibitem{Cheng:2015cca}
Cheng H~Y and Chua C~K 2015 {\em Phys. Rev.\/} {\bf D92} 096009
  (\textit{Preprint} \eprint{1509.03708})

\bibitem{Chilikin:2014bkk}
Chilikin K {\em et~al.\/} (Belle) 2014 {\em Phys. Rev.\/} {\bf D90} 112009
  (\textit{Preprint} \eprint{1408.6457})

\bibitem{Aaij:2016phn}
Aaij R {\em et~al.\/} (LHCb) 2016 {\em Phys. Rev. Lett.\/} {\bf 117} 082002
  (\textit{Preprint} \eprint{1604.05708})

\bibitem{Aubert:2008aa}
Aubert B {\em et~al.\/} (BaBar) 2009 {\em Phys. Rev.\/} {\bf D79} 112001
  (\textit{Preprint} \eprint{0811.0564})

\bibitem{Aaij:2016iev}
Aaij R {\em et~al.\/} (LHCb) 2016 {\em accepted by Phys. Rev. Lett.\/}
  (\textit{Preprint} \eprint{1608.00435})

\bibitem{CMS:2016fvl}
{CMS Collaboration} (CMS) 2016

\bibitem{Aaij:2014asa}
Aaij R {\em et~al.\/} (LHCb) 2014 {\em Phys. Rev. Lett.\/} {\bf 113} 152003
  (\textit{Preprint} \eprint{1408.0971})

\bibitem{Aaij:2014gka}
Aaij R {\em et~al.\/} (LHCb) 2015 {\em Phys. Lett.\/} {\bf B742} 29--37
  (\textit{Preprint} \eprint{1411.6899})

\bibitem{Aaij:2016xxs}
Aaij R {\em et~al.\/} (LHCb) 2016 {\em Phys. Lett.\/} {\bf B759} 313--321
  (\textit{Preprint} \eprint{1603.07037})

\bibitem{Aaij:2016xas}
Aaij R {\em et~al.\/} (LHCb) 2016  (\textit{Preprint} \eprint{1607.06134})

\bibitem{DescotesGenon:2009ja}
Descotes-Genon S, He J, Kou E and Robbe P 2009 {\em Phys. Rev.\/} {\bf D80}
  114031 (\textit{Preprint} \eprint{0907.2256})

\end{thebibliography}

\end{document}